\documentclass[a4paper,11pt]{article}
\usepackage{pos}

\pdfoutput=1

\usepackage{graphicx}
\usepackage[capitalise]{cleveref}
\usepackage{subcaption}
\usepackage{svg}
 \usepackage{tabularx, booktabs, makecell}

\usepackage{float}
\usepackage{xcolor}

 \newcolumntype{Y}{>{\centering\arraybackslash}X}
 
\begin{document}
 
\title{
\vspace*{-0.75cm}

\begin{minipage}{\textwidth}
\begin{flushright}

\end{flushright}
\end{minipage}\\[15pt]

\vspace*{+0.75cm}
        Fourier Accelerated HMC in the $SU(N)\times SU(N)$ Principal Chiral Model}

\ShortTitle{Fourier Accelerated HMC in the $SU(N)\times SU(N)$ Principal Chiral Model}

\author[a]{Roger~Horsley}
\author[a,b]{Pablo~Morande}
\author*[a]{Brian~Pendleton}

\affiliation[a]{School of Physics and Astronomy, University of Edinburgh,
                Edinburgh, EH9 3FD, UK}
\affiliation[b]{Department of Applied Mathematics and Theoretical Physics, Centre for Mathematical Sciences,
University of Cambridge, Wilberforce Road, Cambridge, CB3 0WA, UK}

\emailAdd{rhorsley@ph.ed.ac.uk}
\emailAdd{pm843@cam.ac.uk}
\emailAdd{bjp@ph.ed.ac.uk}

\abstract{We report the results of Fourier-accelerated HMC simulations of 2D $SU(N) \times SU(N)$ principal chiral
models for N = 2, 3, 4, 6, 9. These models share several key properties with 4D QCD, such as asymptotic freedom and dynamical mass generation. Even for modest correlation lengths, we
find integrated autocorrelation times are decreased by an order of magnitude relative to standard
HMC, with relatively little computational overhead. Our results also suggest that the relative advantage of Fourier Acceleration over traditional HMC decreases as N increases, possibly due to the
enlarged group space associated with larger N. Our Monte Carlo results agree with the mass
spectra and continuum scaling behaviour predicted by the exact solution obtained using the Bethe
ansatz.}

\FullConference{%
   The 41st International Symposium on Lattice Field Theory (LATTICE2024)\\
   28 July - 3 August 2024\\
   Liverpool, UK\\}

\maketitle
\section{Introduction}
\noindent
In lattice simulations, critical slowing down refers to the significant increase in run time required to obtain results as the lattice spacing, $a$, is reduced. This phenomenon arises because our algorithms generate Monte Carlo configurations which become highly correlated in the $a\to0$ limit. In practice, the degree of autocorrelation in the configurations is measured through the integrated autocorrelation time (IAT), $\tau_{IAT}$, which can be interpreted as the number of configurations between independent samples. In the case of Hybrid Monte Carlo (HMC) \cite{Duane:1987de}, critical slowing down emerges because the algorithm treats both long- and short-wavelength modes equally. The short wavelength modes evolve faster in the simulation and set a strict upper limit on the integration step size for accurate integration. This means the more physical long wavelength modes, which evolve more slowly, barely evolve in each molecular dynamics step, resulting in highly correlated configurations. 

\vspace{2mm}
\noindent
Consider a free theory, the equations of motion lead to the dispersion relation $\omega^2(p) = p^2 + m^2$ \cite{Duane:1986fy}, which shows that modes with higher (shorter) momentum (wavelength) evolve faster than those with lower (longer) ones. Generally, the maximum integration step size $dt$ is inversely proportional to the maximum possible frequency \cite{Duane:1986fy}. Therefore, the number of steps required for the mode with the lowest momentum to complete a whole cycle is given by:
\begin{equation}
    \frac{T}{dt} = \frac{\omega(p_{max})}{\omega(p_{min})} =\sqrt{\frac{p^2_{max} +m^2}{p^2_{min}+m^2}}
    \label{eq:ratio}
\end{equation}
Critical slowing down occurs when this ratio diverges as $a\to 0$ as more modes are included in the simulation. One possible way to circumvent this issue is Fourier Acceleration (FA), in which the dynamics are modified such that the slow modes are accelerated and the fast ones are slowed down by adding a momentum-dependent "mass" to the system. Ideally, this would make the evolution speed of the modes independent of the momentum. To achieve this, the Hamiltonian is modified by introducing the inverse kernel of the action in the distribution of the momenta.

\vspace{2mm}
\noindent
Returning to the free theory, this modification yields a new dispersion relation of $\omega^2(p) = \frac{p^2+m^2}{p^2+M^2}$ \cite{Duane:1986fy} where $M$ is a tunable parameter included in the kernel introduced in the momenta distribution. Choosing $M = m$ makes the evolution rate independent of the mode's momentum, completely avoiding critical slowing down. While being exemplified through the free theory, the idea of modifying the dynamics of the system to change the evolution rates of the different modes can, in principle, be applied to asymptotically free theories (like QCD) \cite{Sheta:2021hsd,Huo:2024lns,Jung:2024nuv} in which the problematic high momentum modes enter the simulation as asymptotically free modes.
\noindent

\section{The SU(N) x SU(N) Principal Chiral Model}
\noindent
In the continuum, the 2D $SU(N) \times SU(N)$ model is defined via the Euclidean Lagrangian:
\begin{align}
    \mathcal{L} = \frac{1}{T}\,\operatorname{Tr}\,{\partial_\mu U\partial_\mu U^\dag}&& U\in SU(N)
\end{align}
On the lattice, the model is defined via the action:
\begin{align}
     S = -\beta N\sum_{x,\mu>0}\operatorname{Tr}\{U^\dag_xU_{x+\mu}+U_{x+\mu}^\dag U_x\}   && \beta = \frac{1}{NT}
\end{align}
\noindent 
This model shares properties with QCD such as asymptotic freedom and dynamical mass generation. Still, it only possesses a $SU(N)\times SU(N)$ global symmetry and is not locally gauge invariant. Furthermore, it is possible to obtain analytical predictions using the Bethe ansatz to calculate the exact solution \cite{Balog:1992cm}. By studying the effect of Fourier acceleration in this model, we aim to establish a foundation for investigating more complex theories and, ultimately, QCD.

\subsection{HMC \& Fourier Acceleration in the Principal Chiral Model}
\noindent The field elements $U_x$ belong to the $SU(N)$ group and hence they can be expressed as:
\begin{equation}
    U_x = \exp\{i\alpha_x^a\lambda_a\}
\end{equation}
Where $\lambda_a$ are a generalization of the Gell-Mann matrices. We define the canonical momentum $\pi^a$ conjugate to the $\alpha_x^a$ and we choose:
\begin{equation}
    H = \frac{1}{2}\sum_x \pi_x^a\pi_x^a + S
\end{equation}
The equations of motion are:
\begin{align}
    \dot U_x = i\pi_xU_x && \pi_x = \pi_x^a \lambda_a 
\end{align}
\begin{equation}
     \dot\pi_x=-2i\beta N\left[ \sum_{\mu>0} \left(\left(U_{x+\mu} +U_{x-\mu}\right)U_x^\dag - h.c\right) -  \frac{1}{N}\operatorname{Tr}\{...\}I\right]
\end{equation}
The Leapfrog discretisations are:
\begin{align}
    \pi_x\left(t+\frac{dt}{2}\right) = \pi_x\left(t-\frac{dt}{2}\right) + dt\dot \pi_x\left(t\right) &&
    U_x(t+dt) = \exp\left\{i\pi_x\left(t+\frac{dt}{2}\right)dt\right\}U_x(t)
\end{align}
In order to implement FA we introduce the inverse kernel of the action in the canonical momentum distribution, that is:
\begin{align}
    H_{FA} = \frac{1}{2}\sum_{x,y} \pi_x K^{-1}_{x,y}(M)\pi_y + S && S = \beta N\sum_{x,y}\operatorname{Re}\operatorname{Tr}\{U^\dag_x K_{x,y}(0)U_y\}   
\end{align}
In Fourier space the kernel is diagonal, simplifying its inversion. The Fourier transform and its inverse on an $L^D$ lattice are defined as
\begin{align}
     \Tilde{f}(k)=\mathcal{F}[f(x)]_k = \sum_x f(x) \exp\left\{-2\pi i\frac{k\cdot x}{L}\right\} \\ f(x)=\frac{1}{L^D}\mathcal{F}^{-1}[\Tilde{f}(k)]_x = \sum_k f(k) \exp\left\{2\pi i\frac{k\cdot x}{L}\right\}
\end{align}
The inverted kernel in Fourier space is given by:
\begin{equation}
    \Tilde{K}^{-1}_{k,k'}(M) = \delta_{k,k'}\frac{1}{\sum_\mu 4\sin^2\left(\frac{\pi k_\mu}{L}\right)+M^2}
\end{equation}
This modification leaves the momentum's time evolution unchanged but complicates the equation for the field. In discrete time this becomes:
\begin{equation}
         U_x(t+dt) = \exp\left\{i\mathcal{F}^{-1}\left[\Tilde{K}^{-1}_k(M)\Tilde{\pi}_k\left(t+\frac{dt}{2}\right)\right]_x dt\right\}U_x(t)
\end{equation}
For details on the implementation of HMC and FA in this theory see \cite{Horsley:2023fhj}.
\section{Results}
\subsection{Acceleration}
\begin{figure}[H]
\centering

\begin{subfigure}{0.45\textwidth}
  \centering
  \def\svgwidth{1.09\columnwidth}
  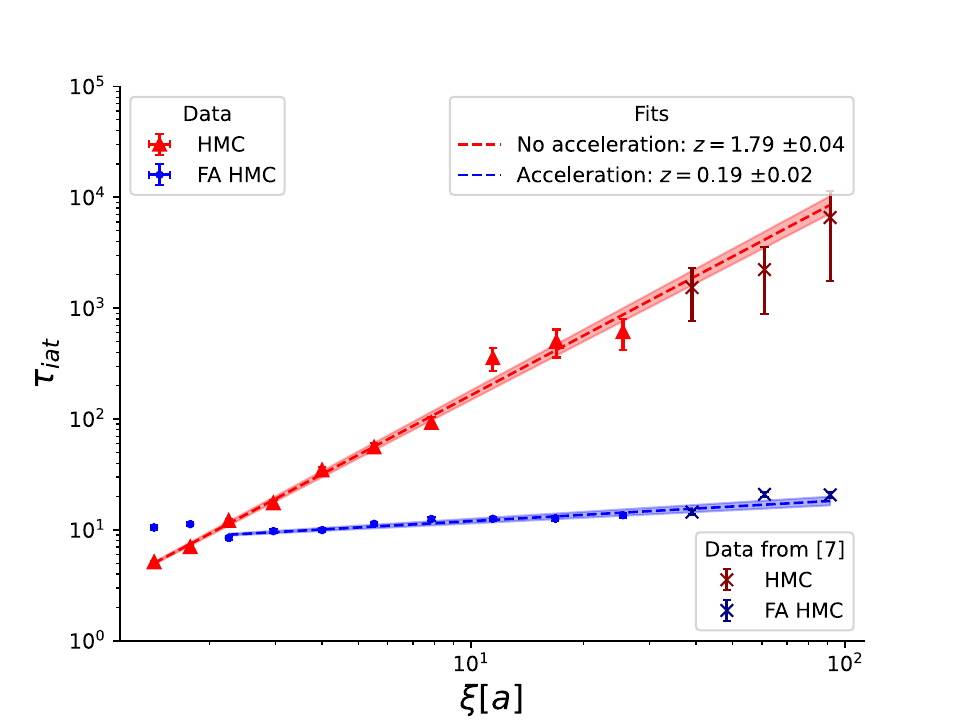
  \subcaption{$SU(2)$}
\end{subfigure}\begin{subfigure}{0.45\textwidth}
  \centering
  \def\svgwidth{1.09\columnwidth}
    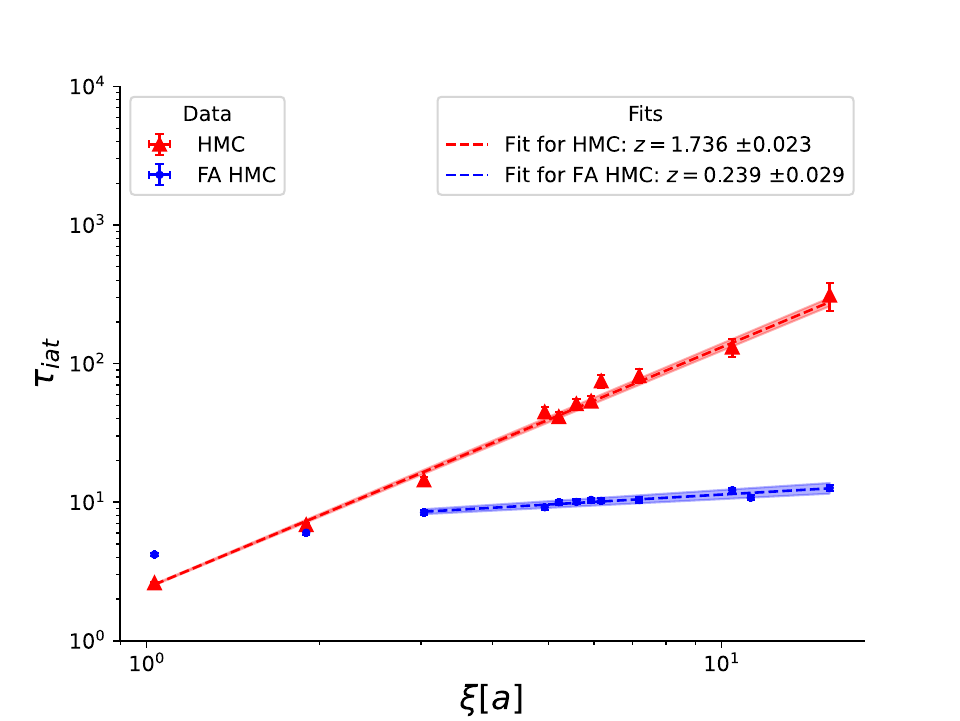
    \subcaption{$SU(3)$}

\end{subfigure}%

\begin{subfigure}{0.45\textwidth}
  \centering
  \def\svgwidth{1.09\columnwidth}
  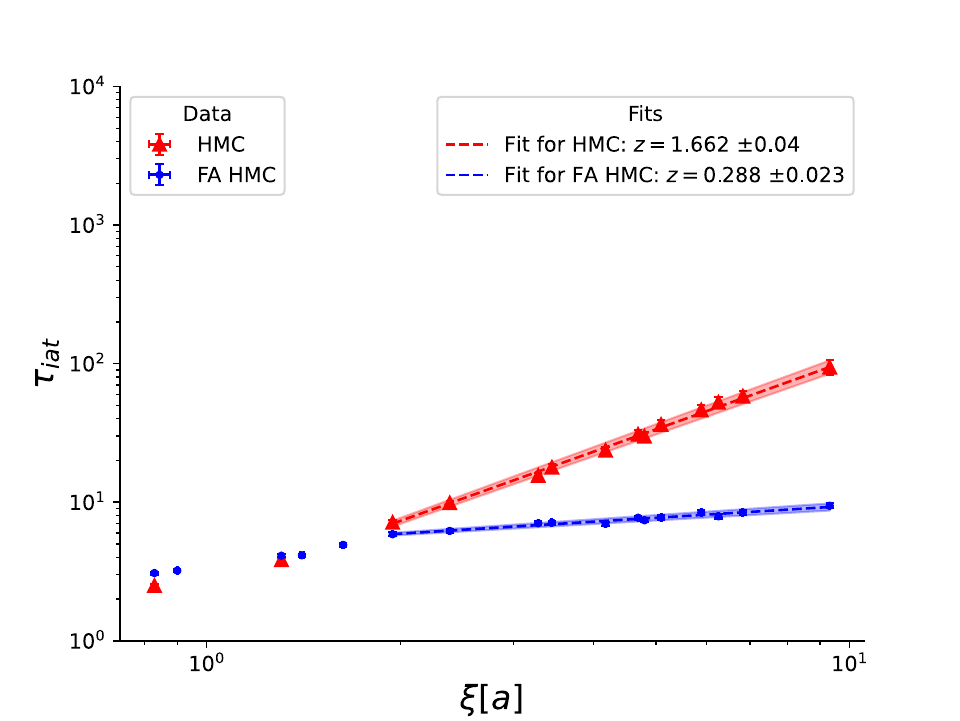
   \subcaption{$SU(4)$}
\end{subfigure}%
\begin{subfigure}{0.45\textwidth}
  \centering
  \def\svgwidth{1.09\columnwidth}
  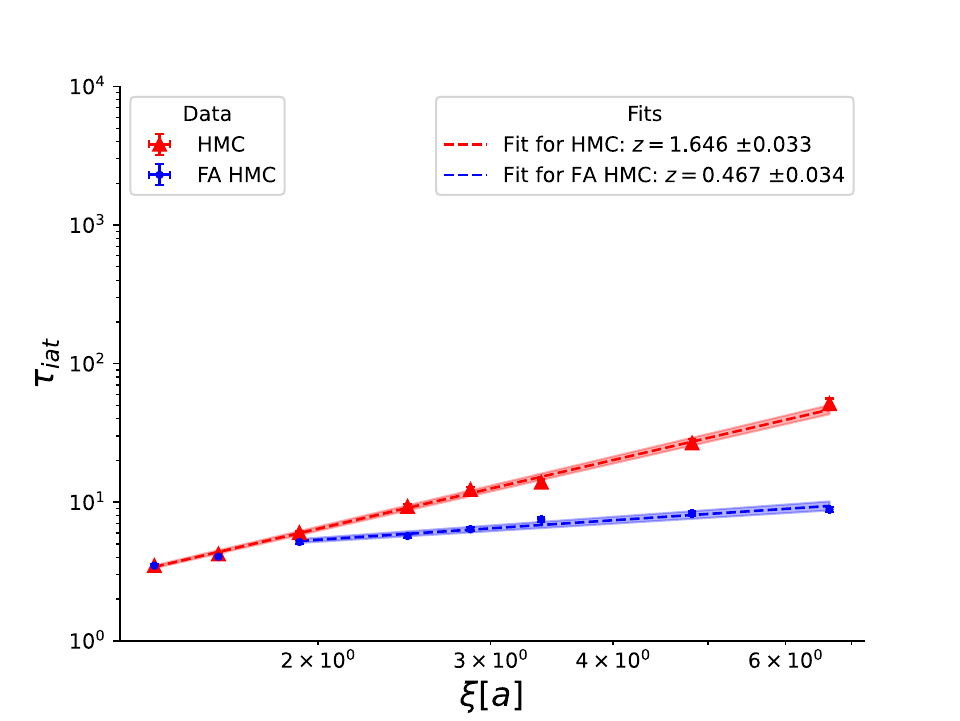
  \subcaption{$SU(6)$}
\end{subfigure}%
 \caption{The plots show the autocorrelation time $\tau_{IAT}$ as a function of the correlation length for several $SU(N)$ in HMC and FA HMC.}
\label{fig:superiat}

%\begin{subfigure}{0.45\textwidth}
%    \centering
%   \includesvg[scale=0.45]{Susceptibility_iat_SU_=_2_Order_=_0_N_Order_=_0_N_measurements_=_100000_N_Thermal_=_10000.svg}
%    \subcaption{$SU(2)$}
%
%\end{subfigure}\begin{subfigure}{0.45\textwidth}
%
%\begin{subfigure}{0.45\textwidth}
%  \centering
%  \def\svgwidth{1.09\columnwidth}
%   \input{Susceptibility_iat_SU_=_2_Order_=_0_N_Order_=_0_N_measurements_=_100000_N_Thermal_=_10000_svg-tex.pdf_tex}
%    \subcaption{$SU(2)$}
%%
%\end{subfigure}\begin{subfigure}{0.45\textwidth}
%    \centering
%    \includesvg[scale=0.45]{Susceptibility_iat_SU_3_Order_10_N_Order_10_N_measurements_100000_N_Thermal_10000.svg}
%    \subcaption{$SU(3)$}
%
%\end{subfigure}%
%
%\begin{subfigure}{0.45\textwidth}
%    \centering
%   \includesvg[scale=0.45]{Susceptibility_iat_SU_4_Order_10_N_Order_10_N_measurements_100000_N_Thermal_10000.svg}
%    \subcaption{$SU(4)$}
%\end{subfigure}%
%\begin{subfigure}{0.45\textwidth}
%    \centering
%    \includesvg[scale=0.45]{Susceptibility_iat_SU_6_Order_10_N_Order_10_N_measurements_100000_N_Thermal_10000.svg}
%    \subcaption{$SU(6)$}
%\end{subfigure}%
% \caption{The plots show the autocorrelation time $\tau_{IAT}$ as a function of the correlation length for several $SU(N)$ in HMC and FA HMC.}
%\label{fig:superiat}

\end{figure}

\noindent
To assess the algorithm's efficiency relative to traditional HMC, we measured the IAT of the model's susceptibility, a long-wavelength observable prone to critical slowdown. Following \cite{madras,stats}, the IAT and its error were calculated for $N = 2, 3, 4, 6$ at various correlation lengths using both FA and traditional HMC. The data were fitted to the ansatz $\tau_{IAT} = b\xi^z$, where $z$ is the dynamical critical exponent and $\xi = \frac{1}{m_0a}$ is the system's correlation length (inverse of the lightest mass). This ansatz captures the correct behaviour for the IAT for sufficiently large correlation lengths and it was found to describe the data well. The acceleration mass parameter $M$ was chosen to be $M = \xi^{-1}$ for each point, this was found to yield a near to optimal acceleration rate using \cref{eq:cost} to compare different runs. Details of $\xi$ calculations for each $\beta$-value are provided in \cref{sec:mass}. To ensure consistency, only runs with acceptance rates near $0.7$ were included. Results are shown in \cref{fig:superiat}, with red and blue points representing HMC and FA HMC measurements, respectively.

\noindent
The reduction in the critical exponent $z$ demonstrates that FA HMC outperforms traditional HMC, though it does not fully eliminate critical slowing down. For $N = 3, 4, 6$, with correlation lengths up to 10, FA achieved a tenfold speedup. For $N = 2$, the final three data points from \cite{Horsley:2023fhj} align well with our extrapolated results, and suggest a two-orders-of-magnitude improvement at larger correlation lengths. Notably, our $z = 0.24 \pm 0.03$ for the $SU(3)$ model is significantly lower than $z = 0.45 \pm 0.02$ reported in \cite{Mana:1996pk} using Multigrid Monte Carlo, highlighting FA HMC's success when compared to other methods used to reduce autocorrelations. However, FA HMC's relative advantage over HMC decreases with increasing $N$, possibly due to the larger group space at higher $N$.

\vspace{2mm}

\begin{figure}[H]
\centering
\begin{subfigure}{0.32\columnwidth}
\centering
  \def\svgwidth{1.09\columnwidth}
       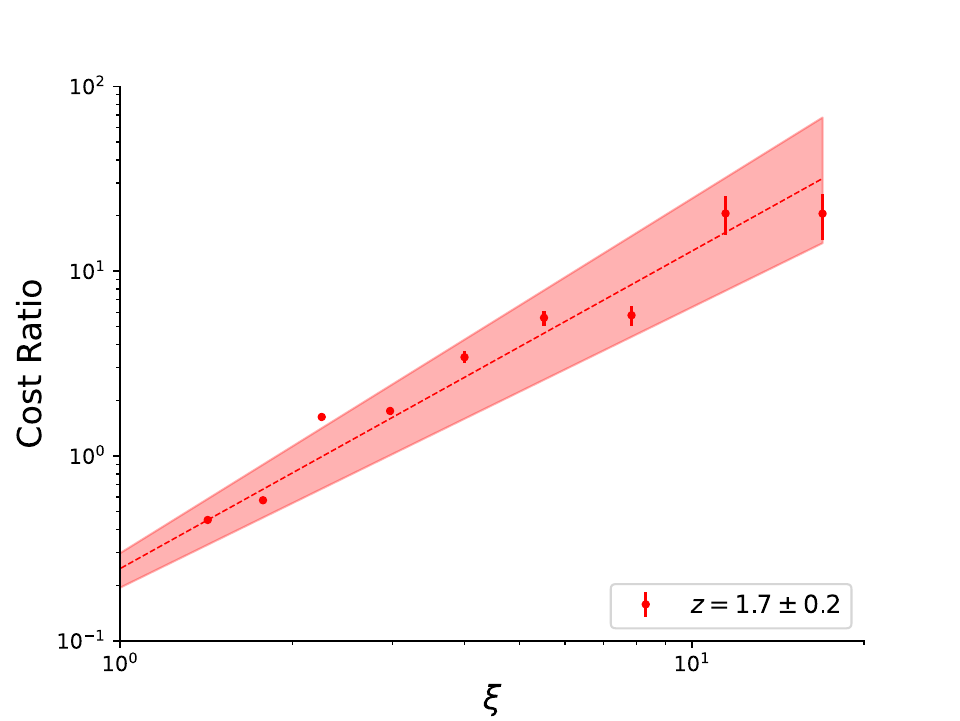
       \subcaption{$SU(2)$}
\end{subfigure}
\begin{subfigure}{0.32\columnwidth}
\centering
  \def\svgwidth{1.09\columnwidth}
       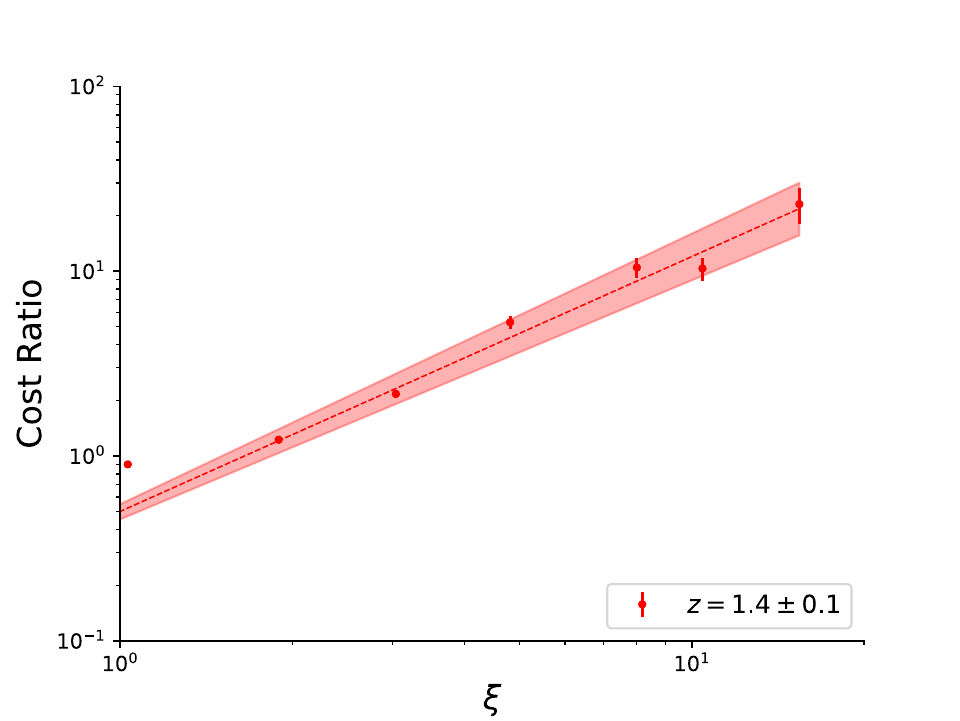
       \subcaption{$SU(3)$}
\end{subfigure}
\begin{subfigure}{0.32\columnwidth}
\centering
  \def\svgwidth{1.09\columnwidth}
       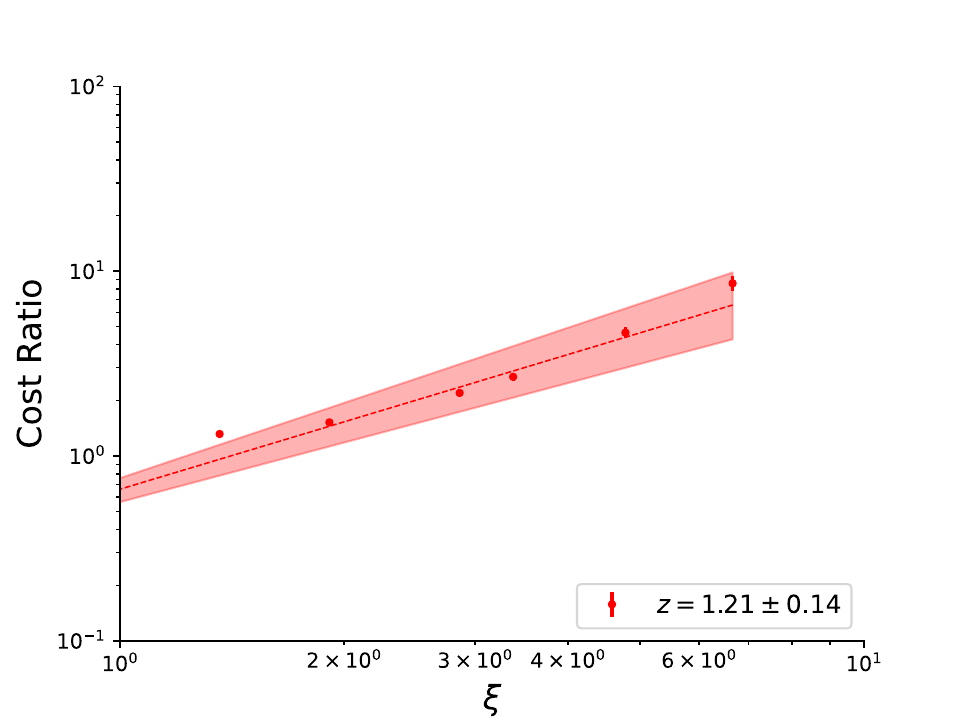
       \subcaption{$SU(6)$}
\end{subfigure}

 \caption{Cost ratio as a function of the correlation length for several $SU(N)$.}
\label{fig:supercost}
\end{figure}

\noindent
Fourier acceleration introduces a more complex molecular dynamics evolution of the field and conjugate momentum. Ensuring that the IAT's reduction compensates for any potential increase in the algorithm's runtime is crucial. To measure this, we define the cost of a sampling algorithm as:
\begin{equation}
    \text{Cost} = \frac{\text{Computer Time}}{\text{Nº Effective Configurations}}=\frac{\text{Computer Time}}{\text{Nº Configurations}} \: \tau_{IAT}
    \label{eq:cost}
\end{equation}
\noindent
The plots in \cref{fig:supercost} show the traditional HMC and FA HMC cost ratio for various $\xi$ and N values. We followed a similar procedure to the one outlined above (acceptance $\approx 70\%$) to select the points used for the fit to the ansatz with the additional criteria of only showing the points in which $L\gtrsim 8.5\xi$ (to reduce finite volume effects). The results confirm that FA HMC outperforms HMC, achieving an order of magnitude speed-up for moderate $\xi \sim 10$ even when accounting for run-time. Once more, the advantage of FA HMC diminishes as $N$ increases.

%\begin{figure}[H]
%\centering
%\begin{subfigure}{0.32\columnwidth}
%\centering
%       \includesvg[scale = 0.32]{cost_ratio_SU_2_Order_0_N_Order_0_N_measurements_10000_N_Thermal_100000.svg}
%       \subcaption{$SU(2)$}
%\end{subfigure}
%\begin{subfigure}{0.32\columnwidth}
%\centering
%       \includesvg[scale = 0.32]{cost_ratio_SU_3_Order_10_N_Order_10_N_measurements_100000_N_Thermal_10000.svg}
%       \subcaption{$SU(3)$}
%\end{subfigure}
%\begin{subfigure}{0.32\columnwidth}
%\centering
%       \includesvg[scale = 0.32]{cost_ratio_SU_6_Order_10_N_Order_10_N_measurements_100000_N_Thermal_10000.svg}
%       \subcaption{$SU(6)$}
%\end{subfigure}
%
%
% \caption{Cost ratio as a function of the correlation length for several $SU(N)$.}
%\label{fig:supercost}
%\end{figure}

\noindent
\subsection{Mass Spectrum}
\label{sec:mass}
\noindent The extraction of the correlation length $\xi$ for a given $\beta$ is crucial to study the variation the $\tau_{IAT}$ as a function of $\xi$. To extract the mass of the lightest state (inverse of correlation length) in lattice units we make use of (zero momentum) two-point functions:
\begin{equation}
     C_{ww}(t) = \frac{1}{V}\left\langle\sum_{x,y,\tau} 2\operatorname{Re}\operatorname{Tr}(U_{x,\tau}U^\dag_{y,\tau+t})\right\rangle = \sum_i A_i \cosh\left(m_i\left(t-\frac{L}{2}\right)\right)
     \label{eq:C_ww}
\end{equation}
which, given the periodic boundary conditions, is expected to have a spectral decomposition given by the second equality of \cref{eq:C_ww} where the lowest energy state dominates as long as the measurements are taken at large enough time. This provides a reliable method to extract the lightest mass $m_0$. To determine $m_0$, we used the effective mass $m_{eff}$:
\begin{align}
     m_{eff}(t) =  \cosh^{-1}\left(\frac{C_{ww}(t+1) + C_{ww}(t-1)}{2C_{ww}(t)}\right)
\end{align}
\noindent
The effective mass was fitted to a constant ($m_0$) using a version of the Akaike Information Criterion (AIC) model averaging procedure \cite{Segner:2023igh} to mitigate the bias in selecting the fitting region. 
This approach assigned weights to fits based on their chi-squared values and the extent of their fitting ranges. Correlation lengths were determined for $N = 2, 3, 4, 6, 9$ across several beta-values, showing good agreement with literature results \cite{Rossi:1994yp} but with much smaller error bars. Below, we present findings for SU(3).

 \begin{table}[H]
\begin{tabularx}{\textwidth}{YYYY}  \toprule
$\beta$ &L&$\xi$ & $\xi$ Ref \cite{Rossi:1994yp} \\ 
\midrule
$0.29$ &$82$& $8.02(2)$ & $8.08(13)$ \\
$0.3$ & $90$  & $10.44(2)$& $10.48(12)$ \\
$0.315$ & $120$  &$15.42(3)$ &$15.8(4)$ \\          \bottomrule
\end{tabularx}
\caption{Correlation Length values obtained for the SU(3) model.}
\label{table:SU3}
\end{table}
\noindent
The exact solution in the continuum \cite{Balog:1992cm} predicts only one bound state for $SU(2)$, implying a constant $m_{eff}$ even at short time separations. However, significant deviations from this behaviour were found in the $SU(2)$ results at short times. Since the continuum model does not contain excited states, the physical interpretation of this contamination is less clear. Nevertheless, a careful examination of this phenomenon revealed that it is likely a lattice artifact as the physical distance over which it occurs shrinks to zero in the continuum limit.

\vspace{2mm}
\noindent
This analysis can be extended to excited energy levels in the spectrum by replacing the field $U$ with an operator $O$ in the definition of $C_{ww}$, which overlaps strongly with the state of interest. Then, the modified two-point correlation function will present (ideally) only that state in the spectral decomposition (allowing for contamination from excited states). The relevant operators for this theory are given in \cite{Rossi:1994yp}, with the first excited state operator being:
\begin{equation}
    O_{abcd} = U_{ab}U_{cd} - U_{ad}U_{cb}
\end{equation}
We extracted the first excited state mass for $N = 3, 4, 6$. Here, we present the ratio of the lightest state to the first excited state for $N = 4$ alongside the continuum prediction. The ratio approaches the continuum value as $\beta$ increases and the results agree with \cite{Rossi:1994yp} with smaller error bars.

    \begin{table}[H]
\begin{tabularx}{\textwidth}{YYY}  \toprule
$\beta$&L &$\frac{m_1}{m_0}$ \\ \midrule
$0.225$& $24$& $1.509(7)$\\
$0.29$& $82$& $1.439(5)$\\
$0.31$ & $100$ &$1.421(9)$ \\
Continuum && $\sqrt{2} \approx 1.41421356$\\
          \bottomrule
\end{tabularx}

\caption{Ratio between the ground state mass $m_0$ and the first excited state mass $m_1$ for the $SU(4)$ model.}
\label{table:SU42}
\end{table}
\noindent 
\subsection{Asymptotic Scaling}
\begin{figure}
\centering
\begin{subfigure}{0.4\textwidth}
    \centering
  \def\svgwidth{1.08\columnwidth}
  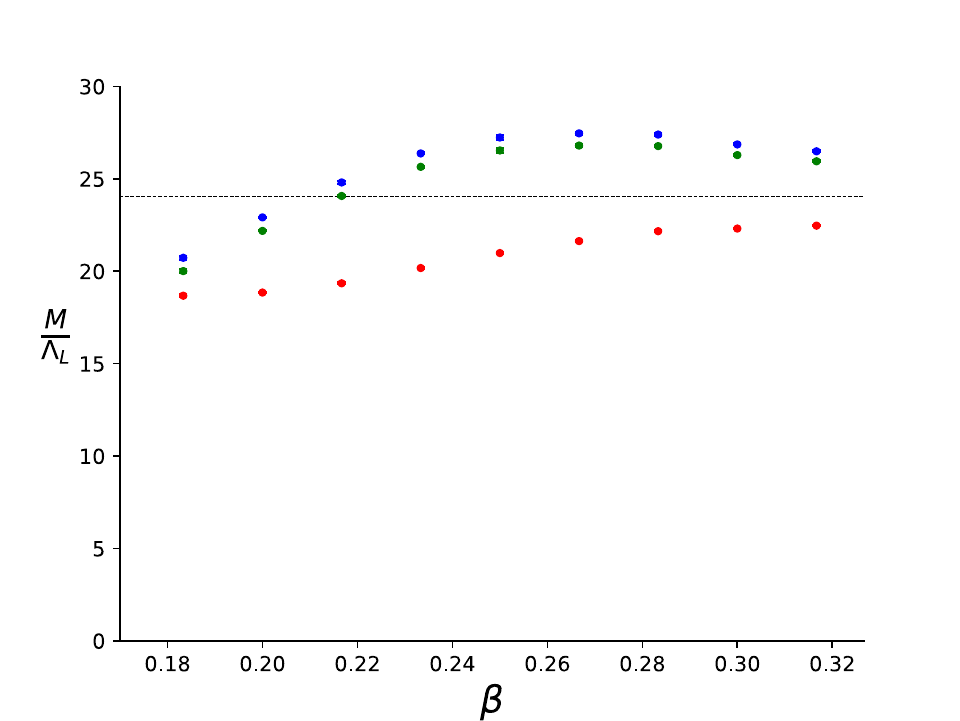
    \subcaption{$SU(2)$}
    \label{fig:As2}
\end{subfigure}%
\begin{subfigure}{0.4\textwidth}
    \centering
  \def\svgwidth{1.02\columnwidth}
    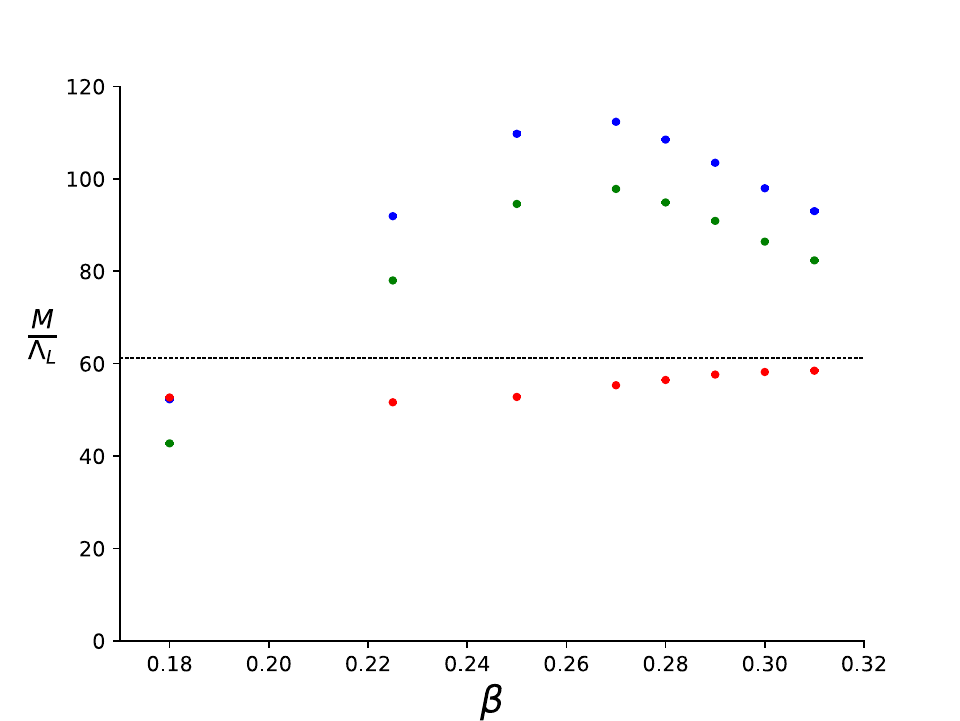
    \subcaption{$SU(4)$}
\end{subfigure}
\begin{subfigure}{0.4\textwidth}
    \centering
  \def\svgwidth{1.08\columnwidth}
    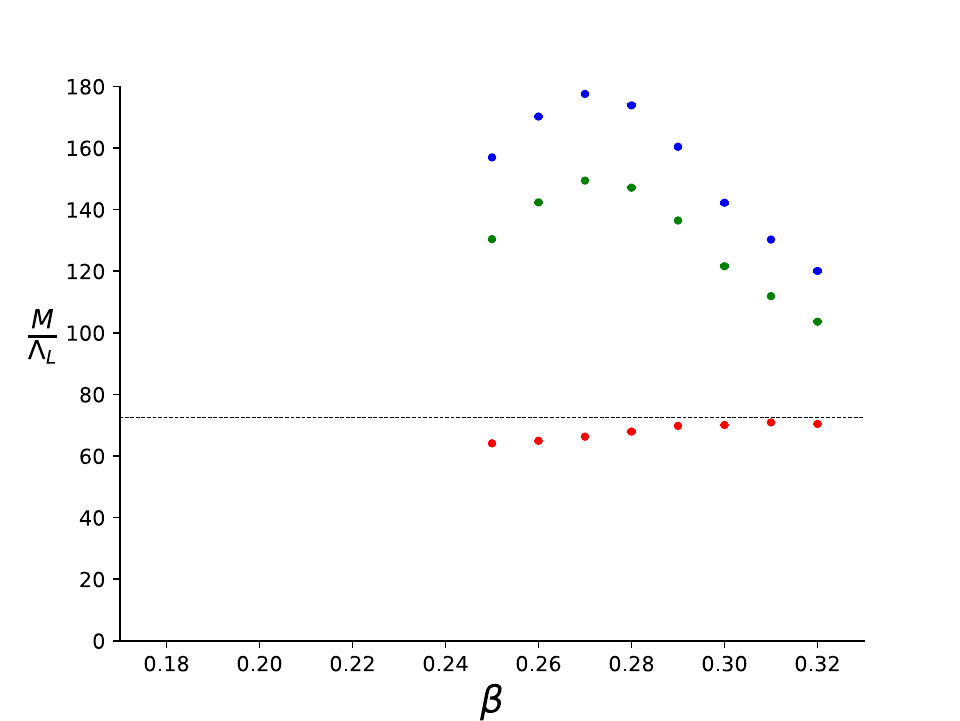
    \subcaption{$SU(6)$}
    \label{fig:As9}
\end{subfigure}%
\begin{subfigure}{0.4\textwidth}
    \centering
  \def\svgwidth{1.08\columnwidth}
   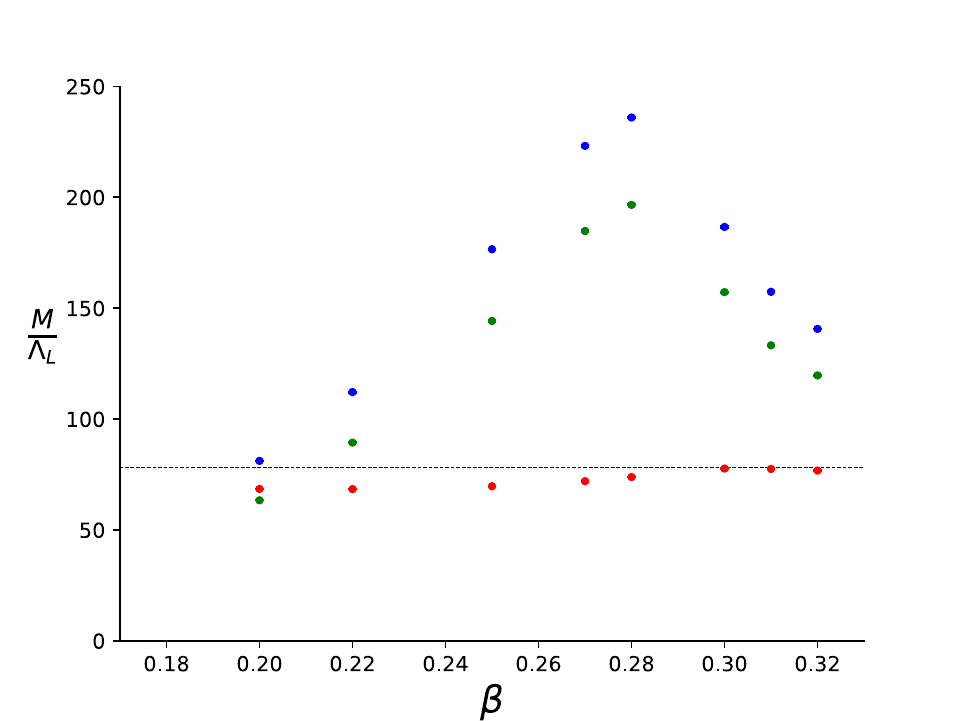
    \subcaption{$SU(9)$}
    \label{fig:As92}
\end{subfigure}

\caption[short]{The plots display the asymptotic scaling for various $N$. Blue and green points represent the Mass-to-$\Lambda$ ratio in the regular scheme at two-loop and three-loop orders, respectively. Red points and the dashed line correspond to the ratio in the energy scheme and the continuum prediction, respectively.}
\label{fig:supermL}
\end{figure}

\noindent
To study rigorously how the continuum limit is approached in this theory and to test how well it is approached in our simulation we look at the lattice beta function, which is expanded in a power series in $T$ %\cite 
:
\begin{align}
   \beta_L(T)\equiv -\frac{\partial T}{\partial \ln a}=- \beta_0 T^2 -\beta_1T^3 + \mathcal{O}(T^4) &&
    \beta_0 = \frac{N}{8\pi} &&
    \beta_1 =\frac{N^2}{128\pi^2}
    \label{eq:beta}
\end{align}
The parameters $\beta_0,\beta_1$ are determined by two-loop perturbation theory \cite{Rossi:1994yp}. Integrating \cref{eq:beta} yields the two-loop relation between $a$ and $T$:
\begin{equation}
    a\Lambda_{L,2l} =\sqrt{\frac{8\pi}{NT}}\exp\left\{-\frac{8\pi}{NT}\right\}
    \label{eq:aT}
\end{equation}
where $\Lambda_{L,2l}$ is an integration constant ($\Lambda-$parameter). \cref{eq:aT} implies that the continuum limit is obtained by taking $\beta\to\infty$, which is the lattice version of asymptotic freedom. It is possible to test how well the continuum limit is approached by ensuring the ratio between physical masses and the $\Lambda$-parameter remains constant (asymptotic scaling). Using the Bethe ansatz solution, the exact ratio between physical masses and the $\Lambda$-parameter in the $\overline{MS}$ scheme can be calculated, and it also allows for determining the ratio between the $\Lambda$-parameters in the lattice and $\overline{MS}$ schemes \cite{Balog:1992cm}:
\begin{equation}
    \frac{m_0}{\Lambda_{L}} = \sqrt{\frac{256N^2}{e\pi}}\sin\left(\frac{\pi}{N}\right)\exp\left\{\pi\frac{N^2-2}{2N^2}\right\}
    \label{eq:MoverLcont}
\end{equation}
Hence, we can test the prediction of \cref{eq:MoverLcont}, and therefore asymptotic freedom, by inspecting the ratio $\frac{m_0}{\Lambda_{L,2l}} = \frac{1}{\xi} \sqrt{8\pi\beta}\exp\{-8\pi\beta\}$ to confirm that it converges to the correct value as $\beta\to\infty$. Our results for several values of $N$ are shown in \cref{fig:supermL} (blue points). These results show a significant overshoot in the $m_0$ over $\Lambda_L$ ratio caused by a dip in the $\beta$ function, which becomes more pronounced as $N$ increases. This feature occurs in the same region where the model's heat capacity peaks (which seems to become singular as $N\to \infty$, indicating a phase transition), \cite{Campostrini:1994ih} suggesting a connection between the two phenomena. In \cite{Rossi:1994yp}, the energy scheme is introduced to fill the dip in the $\beta$-function. The coupling $T$ is redefined in terms of the energy density $e$:
\begin{align}
    e(\beta) = 1+\frac{1}{2DVN^2}\left\langle \frac{S}{\beta}\right\rangle &&  T_E(\beta) = \frac{8N}{N^2-1}e(\beta) && \beta_E = \frac{1}{NT_E(\beta)} 
\end{align}
The same expression is obtained for the $\Lambda$-parameter in this scheme up to two loops ($\Lambda_{E,2l}$). To test asymptotic scaling in this scheme, we look at the ratio defined by:
\begin{equation}
    \frac{m_0}{ \Lambda_{L,2l}}\biggr\rvert_E \equiv \frac{1}{\xi a\Lambda_{E,2l}}\frac{\Lambda_E}{\Lambda_L}
\end{equation}
\noindent
This ratio converges to the same constant given in \cref{eq:MoverLcont} and the results are shown in \cref{fig:supermL} (red points). The energy scheme's success over the regular scheme is evident for $N = 4, 6, 9$ and appears to improve with increasing $N$. We also include the three-loop calculation of the ratio in the regular scheme (green points). Although these corrections improve the results, they are minor, indicating that perturbation theory alone cannot explain the regular scheme's failure or the energy scheme's success \cite{Rossi:1994yp}. In contrast, for $N = 2$, the energy scheme offers no improvement, consistent with \cite{Horsley:2023fhj}. Since this phenomenon is linked to the large $N$ limit (heat capacity becomes singular), the lack of improvement for $N = 2$ might arise from $N$ not being large enough to create a substantial problem. Our results are consistent with \cite{Rossi:1994yp}, with FA allowing greater precision.

%\begin{figure}
%\centering
%\begin{subfigure}{0.4\textwidth}
%    \centering
%  \includesvg[scale=0.4]{M_L_SU_2_Order_0_N_Order_0_N_measurements_100000_N_Thermal_10000_manual_Accel.svg}
%    \subcaption{$SU(2)$}
%    \label{fig:As2}
%\end{subfigure}%
%\begin{subfigure}{0.4\textwidth}
%    \centering
%    \includesvg[scale=0.4]{M_L_SU_4_Order_10_N_Order_10_N_measurements_100000_N_Thermal_10000_manual_Accel.svg}
%    \subcaption{$SU(4)$}
%\end{subfigure}
%\begin{subfigure}{0.4\textwidth}
%    \centering
%    \includesvg[scale=0.4]{M_L_SU_6_Order_10_N_Order_10_N_measurements_100000_N_Thermal_10000_manual_Accel.svg}
%    \subcaption{$SU(6)$}
%    \label{fig:As9}
%\end{subfigure}%
%\begin{subfigure}{0.4\textwidth}
%    \centering
%   \includesvg[scale=0.4]{M_L_SU_9_Order_10_N_Order_10_N_measurements_100000_N_Thermal_10000_manual_Accel.svg}
%    \subcaption{$SU(9)$}
%    \label{fig:As92}
%\end{subfigure}
%
%\caption[short]{The plots display the asymptotic scaling for various $N$. Blue and green points represent the Mass-to-$\Lambda$ ratio in the regular scheme at two-loop and three-loop orders, respectively. Red points and the dashed line correspond to the ratio in the energy scheme and the continuum prediction, respectively.}
%\label{fig:supermL}
%\end{figure}

\section{Summary \& Outlook}
\noindent
We applied Fourier Acceleration to the $SU(N) \times SU(N)$ Chiral Model, demonstrating its ability to mitigate critical slowing down and improve computational efficiency compared to HMC. However, FA's advantage diminishes as $N$ increases, likely due to the enlarged group space. Additionally, we introduced a method for extracting the mass spectrum using the Akaike Information Criterion and analyzed the model's asymptotic scaling. Future work includes extending this approach to locally gauge-invariant theories, such as ongoing studies on 
$U(1)$ gauge theories. Additionally, optimizing algorithm hyperparameters like the molecular dynamics trajectory length (currently fixed to one) alongside the acceleration mass $M$ using our cost definition (\cref{eq:cost}) could further enhance FA’s acceleration rate, as explored in other models \cite{Duane:1988vr,Ostmeyer:2024amv}.

%----------------------------------------------------------------------------
\acknowledgments 
\noindent
RH is supported in part by the STFC grant ST/X000494/1. PM has been partially supported by STFC consolidated grant ST/X000664/1. For the purpose of open access, the authors have applied a Creative Commons Attribution (CC BY) licence to any author-accepted manuscript version arising from this submission.

% ------------------------------------------------------------------

%----------------------------------------------------------------------------

\end{document}